\documentstyle [12pt] {article}
\begin{document}
\setcounter{page}{1}
\def\theequation{\arabic{section}.\arabic{equation}}
\def\theequation{\thesection.\arabic{equation}}
\setcounter{section}{0}

\title{On the solar neutrino problem in \\the relativistic field theory model of the deuteron }

\author{H. Oberhummer\thanks{E--mail: ohu@kph.tuwien.ac.at, Tel.: +43--1--58801--5574, Fax: +43--1--5864203},  A. N. Ivanov\thanks{E--mail: ivanov@kph.tuwien.ac.at, Tel.: +43--1--58801--5598, Fax: +43--1--5864203}~\thanks{Permanent Address:
State Technical University, Department of Theoretical
Physics, 195251 St. Petersburg, Russian Federation}, N.I. Troitskaya\thanks{Permanent Address:
State Technical University, Department of Theoretical
Physics, 195251 St. Petersburg, Russian Federation}, M. Faber\thanks{E--mail: faber@kph.tuwien.ac.at, Tel.: +43--1--58801--5598, Fax: +43--1--5864203}}

\date{}

\maketitle

\begin{center}
{\it Institut f\"ur Kernphysik, Technische Universit\"at Wien, \\ 
Wiedner Hauptstr. 8-10, A-1040 Vienna, Austria}
\end{center}

\vskip1.0truecm
\begin{center}
\begin{abstract}
The relativistic field theory model of the deuteron suggested by us previously is applied to the calculation of the reaction rate of the low--energy two--proton fusion p + p $\to$ D + $e^{\,+}$ + $\nu_{\rm e}$. The theoretical prediction of the reaction rate obtained is 2.9 times  larger than given by the potential approach. This leads to a strong suppression of the high energy solar neutrino fluxes.
\end{abstract}
\end{center}
\vspace{0.2in}

\vskip1.0truecm
\begin{center}
\end{center}

\newpage

\section{Introduction}

\hspace{0.2in} The solar neutrino problem has its origin in the strong discrepancy between predictions of the Standard Solar Model (SSM) [1,2] and the observed solar neutrino fluxes [2].The experimental solar neutrino fluxes turned out substantially less than that predicted by SSM [1--3].

One of the solution of this problem is to lower the temperature in the center of the Sun in comparison to that predicted by the SSM: $T_c = 1.55\times 10^7\,{\rm K}$ [1,2]. Indeed, due to strong dependence of the solar neutrino fluxes on ${\rm T}_c$  just a 20$\div$30$\%$ diminishing of $T_c$ leads to a decrease of the neutrino fluxes by more than in order of magnitude [3].

In order to reduce $T_c$  one can resort to the change of physical and chemical phenomenological inputs which determine the structure of the star [3]. We will not enumerate here all of them, and relegate the reader to Ref.~[3].  We just pick up the temperature associated with the magnitude of the reaction rate of the low--energy two--proton fusion, i.e., p + p $\to$ D + e$^+$ + $\nu_{\rm e}$. Since the solar luminosity should be kept constant, an enhancement of the reaction rate magnitude of the two--proton fusion implies a lower $T_c$.

The low--energy electroweak model and non--relativistic approaches, having been applied to the computation of the contribution of strong interactions to the matrix element of the p + p $\to$ D + W$^+$ transition does not leave a room for the substantial change of the cross section magnitude [1,2,4,5]. As has been noted in Ref.~[3] they differ from the mean value by no more than 3$\,\%$.

Recently we have suggested a phenomenological relativistic field theory model of the deuteron [6,7]. In the relativistic field theory model of the deuteron the strong interactions of particles are described by one--nucleon loop diagrams. In Ref.~[7] we have computed the cross section of the radiative proton--neutron capture n + p $\to$ D + $\gamma$ for thermal neutrons and the cross section of the two--proton fusion p + p $\to$ D + e$^+$ + $\nu_{\rm e}$. In the case of thermal neutron radiative capture our result agrees well with the experimental data and the non--relativistic potential approaches. However, in the case of the two--proton fusion we have found an enhancement of the reaction rate by a factor of 2.9 with respect to the  non--relativistic potential approaches [1,2].

This result is due to our model approach using one--nucleon loop diagrams for the description of strong low--energy interactions of the deuteron to other particles. It is well--known that such fermion--loop diagrams should possess anomalies [8--12].
As usually, these anomalies dominate for the amplitudes of strong low--energy interactions of hadrons [8,13--15]. In the case of the two--proton fusion we encounter the dominance of the anomaly of the AAV one--nucleon loop diagrams, that is the diagrams having two axial--vector and one vector vertices. Such an anomalous contribution produced by vacuum fluctuations of virtual nucleons has completely a quantum field theory nature and cannot be restored within the potential approach describing strong low--energy interactions in terms of the overlap integral of the wave--functions of the protons and the deuteron. The result of the computation of the AAV--anomaly depends on the shift of the virtual nucleon momentum. This is similar to the well--known  Adler--Bell--Jackiw--Bardeen anomaly [8,9] described by the AVV--fermion diagram. This ambiguity can be fixed by the requirement of gauge invariance under gauge transformations of the deuteron field.
This is similar to the removal of the ambiguity appearing for the computation of the AVV--anomaly.

The paper is organized as follows. In Sect.~1 we adduce the computation of the cross section and the reaction rate of the two--proton fusion  p + p $\to$ D + e$^+$ + $\nu_{\rm e}$ [7]. In Sect.~2 we apply the obtained cross section to the discussion of the solar neutrino problem. In the conclusion we discuss the obtained results.

\section{Cross section of two--proton fusion}
\setcounter{equation}{0}

\hspace{0.2in} In the relativistic field theory model of the deuteron the effective Lagrangian describing strong interactions of the deuteron with nucleons reads [6,7]
\begin{eqnarray}\label{label2.1}
&&{\cal L}_{\rm tot}(x) = -\frac{1}{2} D^{\dagger}_{\mu\nu}(x) 
D^{\mu\nu}(x) + M^2_D D^{\dagger}_\mu(x) D^{\mu}(x) - ig_{\rm V} [ \bar{p}(x) \gamma^\mu n^c(x)-\nonumber\\
&&-\bar{n}(x) 
\gamma^\mu p^c(x)] D_\mu (x) -ig_{\rm V} [ \bar{p^c}(x) \gamma^\mu n(x) - \bar{n^c}(x) \gamma^\mu 
p(x) ] D^{\dagger}_\mu (x) +\\
&&  + \bar{p}(x) (i\gamma^\mu \partial_\mu - M_{\rm N}) p(x) + \bar{n}(x) 
(i\gamma^\mu \partial_\mu - M_{\rm N}) n(x),\nonumber
\end{eqnarray}
where $D_{\mu}(x)$, $p(x)$ and $n(x)$ are the interpolating fields of the deuteron, proton and neutron, respectively, $D_{\mu\nu}(x)=\partial_\mu D_\nu(x)-\partial_\nu D_\mu(x)$, $M_{\rm D}= 2M_{\rm N} - \varepsilon_{\rm D}$ is the mass of the deuteron, $M_{\rm N}=938\;{\rm MeV}$ is the mass of the proton and neutron, and $\varepsilon_{\rm D}$ is the binding energy. In our approach $\varepsilon_{\rm D}$ is given by [6,7]
\begin{eqnarray}\label{label2.2}
\varepsilon_{\rm D} = \frac{20}{9}\,\frac{g^2_{\rm V} }{\pi^2} \frac{1}{M^2_{\rm D}}\,\Lambda^3_{\rm D}\, =\,\frac{10}{9}\, Q_{\rm D} \,\Lambda^3_{\rm D}. 
\end{eqnarray}
The phenomenological coupling constant $g_{\rm V}$ is connected with the electric quadrupole moment of the deuteron $Q_{\rm D}$ by the relation
\begin{eqnarray}\label{label2.3}
Q_{\rm D} = \frac{2g^2_{\rm V} }{\pi^2} \frac{1}{M^2_{\rm D}}\,.
\end{eqnarray}
By using the experimental value $(Q_{\rm D})_{\rm exp}=0.286\,{\rm fm^2}$ [16] and $M_{\rm D}\simeq 1876\,{\rm MeV}$ one can estimate the value of $g_{\rm V}$ with $g_{\rm V} = \pm 11.3$. Without loss of generality one can use the positive sign, i.e., $g_{\rm V} = 11.3$ [7]. The quantity $\Lambda_{\rm D}=64.843\;{\rm MeV}$ is the 3--dimensional cut--off in one--nucleon loop diagrams. The magnitude of $\Lambda_{\rm D}$ is obtained by the fit of the binding energy $\varepsilon_{\rm D}=2.225\;{\rm MeV} [16]$. We identify $1/\Lambda_{\rm D}$ with the effective radius of the deuteron [6,7]. At $\Lambda_{\rm D}=64.843\;{\rm MeV}$ we get $r_{\rm D}\,=\,1/\Lambda_{\rm D}\,=\,3.043\,{\rm fm}$ [7], which agrees well with the average value of the deuteron radius, i.e., $<r>\,=\,3.140\,{\rm fm}$ [17].

First, for the computation of the effective Lagrangian describing the low--energy pp--scattering we perform keeping the one--pion--exchange contribution
\begin{eqnarray}\label{label2.4}
{\cal L}^{\,\rm pp}_{\,\rm eff}(x)=-\,\frac{g^2_{\,\pi {\rm NN}}(v)}{2M^2_{\,\pi}}\,[\bar{p}(x)\gamma^{\,5}p(x)]\,[\bar{p}(x)\gamma^{\,5}p(x)]\,.
\end{eqnarray}
Here we have denoted $g_{\,\pi {\rm NN}}(v)=g_{\,\pi {\rm NN}}C(v)$ and $C(v)=\sqrt{2\pi\alpha/v}\exp(-\pi\alpha/v)$, $v$ is a relative velocity of interacting protons, which takes into account the Coulomb repulsion between protons at low energies [4]. 

Applying the Fierz transformation we bring the Lagrangian (\ref{label2.4}) to the form

\parbox{11cm}{\begin{eqnarray*}
&&{\cal L}^{\rm pp}_{\,\rm eff}(x) = \frac{g^2_{\pi {\rm NN}}(v)}{8M^2_{\pi}}\,\Big\{[\bar{p}(x) p^c(x)][\bar{p^{\,c}}(x) p(x)]+[\bar{p}(x)\gamma^5 p^c(x)][\bar{p^c}(x)\gamma^5 p(x)]+\\
&&+[\bar{p}(x)\gamma_{\mu}\gamma^5 p^c(x)][\bar{p^c}(x)\gamma^{\mu}\gamma^5 p(x)]+\frac{1}{2}[\bar{p}(x)\sigma_{\mu\nu}p^c(x)][\bar{p^c}(x)\sigma^{\mu\nu}p(x)]\,\Big\}.
\end{eqnarray*}} \hfill
\parbox{1cm}{\begin{eqnarray}\label{label2.5}
\end{eqnarray}}

\noindent The process p + p $\to$ D + ${\rm e}^+$ + $\nu_{\rm e}$ should run via the intermediate W--boson exchange, i.e., p + p $\to$ D + W$^+$ $\to$ D + ${\rm e}^+$ + $\nu_{\rm e}$. The Lagrangian describing the electroweak interactions of the W--boson with proton, neutron, positron and neutrino reads [18,19]
\begin{eqnarray}\label{label2.6}
{\cal L}^{\rm W}_{\rm int}(x)=-\frac{g_{\rm W}}{2\sqrt{2}}[\bar{p}(x)\gamma^{\mu}(1-g_{\rm A}\gamma^5) n(x)+\bar{\nu}_{\rm e}(x)\gamma^{\mu}(1 - \gamma^5) e(x)]\,W_{\mu}(x)
+{\rm h.c.}.
\end{eqnarray}
Here $g_{\rm W}$ is the electroweak coupling constant connected with the Fermi constant $G_{\rm F}=1.166\time 10^{-5}\,{\rm GeV}^{-2}$ and the W--boson mass $M_{\rm W}$ by the relation [18,19]
\begin{eqnarray}\label{label2.7}
\frac{g^2_{\rm W}}{8M^2_{\rm W}}\,=\,\frac{G_{\rm F}}{\sqrt{2}}\,,
\end{eqnarray}
where $g_{\rm A}=1.260 \pm 0.012$ is the axial--vector coupling constant [16] describing the renormalization of the weak axial--vector hadron current by strong interactions.

The interaction
$[\bar{p}(x)\gamma_{\mu}\gamma^5 p^c(x)][\bar{p^c}(x)\gamma^{\mu}\gamma^5 p(x)]$ gives in our approach the main contribution to the amplitude of the transition p + p $\to$ D + W. The corresponding one--nucleon loop diagrams are depicted in Fig.~1.

The effective Lagrangian defined by the diagrams in Fig.~1 is given by
\begin{eqnarray}\label{label2.8}
&&\int\,d^4x\,{\cal L}_{\rm Fig.1}(x)\,=\nonumber\\
&&=\int\,d^4x\,\int\,\frac{d^4x_1\,d^4k_1}{(2\pi)^4}\,\frac{d^4x_2\,d^4k_2}{(2\pi)^4}\,[\bar{p^c}(x)\gamma_{\, \alpha}\gamma^{\,5}p(x)]\,D^{\dagger}_{\mu}(x_1)\,W^{\dagger}_{\nu}(x_2)\,\times\\
&&\times \, e^{-\,i\,k_1\cdot x_1}\,e^{-\,i\,k_2\cdot x_2}\,e^{\,i\,(k_1\,+\,k_2)\cdot x}\,i\,g_{\rm A}\,\frac{g_{\rm W}}{2\sqrt{2}}\,\frac{g^2_{\pi\rm NN}(v)}{M^{\,2}_{\pi}}\,\frac{g_{\rm V}}{64\pi^2}\,\bar{{\cal J}}^{\alpha\mu\nu}(k_1, k_2; Q)\,,\nonumber
\end{eqnarray}
where
\begin{eqnarray}\label{label2.9}
&&\bar{{\cal J}}^{\alpha\mu\nu}(k_1, k_2;Q)\,=\\
&&=\int\,\frac{d^4k}{\pi^2 i}{\rm tr}\,\Bigg\{\gamma^{\alpha}\gamma^5\frac{1}{M_{\rm N}-\hat{k}-\hat{Q}}\gamma^{\mu}\frac{1}{M_{\rm N}-\hat{k}-\hat{Q}-\hat{k}_1}\gamma^{\nu}\gamma^5\frac{1}{M_{\rm N}-\hat{k}-\hat{Q}-\hat{k}_1-\hat{k}_2}\Bigg\}.\nonumber
\end{eqnarray}
The momentum integral $\bar{{\cal J}}^{\alpha\mu\nu}(k_1, k_2;Q)$ defines the structure of the effective Lagrangian, the 4--vector $Q= a\,k_1 + b\,k_2$ describes an arbitrary shift of the virtual momentum of the nucleons in the diagram in Fig.~1. The parameters $a$ and $b$ are free. It should be emphasized that $a$ and $b$ are not phenomenological parameters of the model, they appear only due to the properties of fermion--loop diagrams under the shift of virtual momenta [8,11,12].

Keeping only the anomalous part of the momentum integral [13--15] and the leading order of the expansion in the powers of the momentum $k_2$ coinciding with the momentum of the leptonic pair that is soft, removing then ambiguities connected with the shift of the virtual momentum by the requirement of the gauge invariance under gauge transformations of the deuteron field,  we arrive at the structure function [7]
\begin{eqnarray}\label{label2.10}
\bar{{\cal J}}^{\alpha\mu\nu}(k_1, k_2;Q) = 3\,(k^{\alpha}_1 g^{\nu\mu} - k^{\nu}_1 g^{\mu\alpha}).
\end{eqnarray}
The effective Lagrangian defined by the structure function (\ref{label2.10}) is given by [7]
\begin{eqnarray}\label{label2.11}
{\cal L}_{\rm Fig.1}(x)=-\,g_{\rm A}\,\frac{g_{\rm W}}{2\sqrt{2}}\,\frac{3g_{\rm V}}{8\pi^2}\,\frac{g^2_{\pi\rm NN}(v)}{8M^2_{\pi}}\,\,W^{\dagger\mu}(x)\,D^{\dagger}_{\mu\nu}(x) \,[\bar{p^c}(x)\gamma^{\nu}\gamma^5 p(x)].
\end{eqnarray}
In turn the effective Lagrangian describing the low--energy process p + p $\to$ D + ${\rm e}^+$ + $\nu_{\rm e}$ reads
\begin{eqnarray}\label{label2.12}
{\cal L}_{\rm eff}(x) = g_{\rm A}\,\frac{G_{\rm F}}{\sqrt{2}}\,\frac{3g_{\rm V}}{8\pi^2}\,\frac{g^2_{\pi\rm NN}(v)}{8M^2_{\pi}}\,j^{\mu}(x)\,
D^{\dagger}_{\mu\nu}(x)\,[\bar{p^c}(x)\gamma^{\nu}\gamma^5 p(x)],
\end{eqnarray}
where $j_{\mu}(x) = \bar{\nu}_{\rm e}(x) \gamma_{\mu}(1-\gamma^5) e(x)$ is the leptonic electroweak current.

In the low--energy limit when the 3--momenta of protons tends to zero the effective Lagrangian (\ref{label2.12}) can be reduced to the expression
\begin{eqnarray}\label{label2.13}
{\cal L}_{\rm eff}(x) = i\,g_{\rm A}\,M_{\rm N}\frac{G_{\rm F}}{\sqrt{2}}\,\frac{3g_{\rm V}}{4\pi^2}\,\frac{g^2_{\pi\rm NN}(v)}{8M^2_{\pi}}\,\,j^{\mu}(x)\,
D^{\dagger}_{\mu}(x)\,[\bar{p^c}(x)\gamma^5 p(x)]\,,
\end{eqnarray}
where we have used the relations

\parbox{11cm}{\begin{eqnarray*}
&&[\bar{p^c}(x)\gamma^{\nu}\gamma^5 p(x)]\to - g^{\nu\,0}[\bar{p^c}(x)\gamma^5 p(x)]\,\\
&&D^{\,\dagger}_{\mu\,0}(x)\to -\,i\,M_{\rm D}D^{\,\dagger}_{\mu}(x)
\end{eqnarray*}} \hfill
\parbox{1cm}{\begin{eqnarray}\label{label2.14}
\end{eqnarray}}

\noindent that are valid in the limit of low 3--momenta of protons and the deuteron. 

The amplitude defined by the effective Lagrangian (\ref{label2.13}) reads
\begin{eqnarray}\label{label2.15}
&&{\cal M}({\rm p} + {\rm p} \to {\rm D} + {\rm e}^+ \nu_{e}) = i\,C(v)\,g_{\rm A} M_{\rm N} \frac{G_{\rm F}}{\sqrt{2}}\,\frac{3g_{\rm V}}{4\pi^2}\,\frac{g^2_{\pi\rm NN}}{8M^2_{\pi}}\,\times\nonumber\\
&&\hspace{1in}\times\, e^{\ast}_{\mu}(Q)\,[\bar{u}(k_{\nu})\gamma^{\mu}(1-\gamma^5)v(k_{\rm e})]\,[\bar{u^c}(p_1) \gamma^5 u(p_2)],
\end{eqnarray}
where $\bar{u}(k_{\nu})$ and $v(k_{\rm e})$ are the Dirac bispinors of the neutrino and positron. For convenience we have separated the factor $C(v)$ describing the Coulomb repulsion.

The amplitude Eq.~(\ref{label2.15}) takes into account only the contribution of the one--pion exchange. Since at low energies the pp--system is in the ${^1}{\rm S}_0$ state with the isospin equal $T=1$, there is the contribution of the pole on the unphysical sheet. Following the computation of the cross section of the radiative neutron--proton capture [7] and normalizing the contribution of the pole on the unphysical sheet to the ${^1}{\rm S}_0$ pp--scattering length $a_{\rm S}$ we get 
\begin{eqnarray}\label{label2.16}
&&{\cal M}({\rm p} + {\rm p} \to {\rm D} + {\rm e}^+ \nu_{e}) = i\,C(v)\,g_{\rm A} M_{\rm N} \frac{G_{\rm F}}{\sqrt{2}}\,\frac{3g_{\rm V}}{4\pi^2}\,e^{\ast}_{\mu}(Q)\,[\bar{u}(k_{\nu})\gamma^{\mu}(1-\gamma^5)v(k_{\rm e})]\times\nonumber\\
&&\hspace{1in}\times \,\frac{g^2_{\pi\rm NN}}{8M^2_{\pi}}\Bigg(1 - \frac{8\sqrt{2}\pi\,M^2_{\pi}}{g^2_{\rm \pi NN}}\,\frac{a_{\rm S}}{M_{\rm N}}\Bigg)\,[\bar{u^c}(p_1) \gamma^5 u(p_2)].
\end{eqnarray}
The experimental value of $a_{\rm S}$ is given by $a_{\rm S} = ( -17.1\pm 0.2)\,{\rm fm}$ [16].

For the correct description of the strong low--energy pp coupling to the deuteron the amplitude Eq.~(\ref{label2.16}) should be unitarized [7]. In our approach we can perform the unitarization by summing up an infinite series of proton--proton bubble diagrams [7]. This gives 
\begin{eqnarray}\label{label2.17}
&&{\cal M}({\rm p} + {\rm p} \to {\rm D} + {\rm e}^+ \nu_{e}) = i\,C(v)\,g_{\rm A} M_{\rm N} \frac{G_{\rm F}}{\sqrt{2}}\,\frac{3g_{\rm V}}{4\pi^2}\,e^{\ast}_{\mu}(Q)\,[\bar{u}(k_{\nu})\gamma^{\mu}(1-\gamma^5)v(k_{\rm e})]\,\times\nonumber\\
&&\times\,\frac{\displaystyle \frac{g^2_{\pi\rm NN}}{8M^2_{\pi}}\Bigg(1 - \frac{8\sqrt{2}\pi\,M^2_{\pi}}{g^2_{\rm \pi NN}}\,\frac{a_{\rm S}}{M_{\rm N}}\Bigg)}{\displaystyle 1 - i\,\frac{g^2_{\pi\rm NN}}{8M^2_{\pi}}\Bigg(1 - \frac{8\sqrt{2}\pi\,M^2_{\pi}}{g^2_{\rm \pi NN}}\,\frac{a_{\rm S}}{M_{\rm N}}\Bigg)\,\frac{M^2_{\rm N}}{2\pi}\,v}\,[\bar{u^c}(p_1) \gamma^5 u(p_2)],
\end{eqnarray}
where $v$ is the relative velocity of the protons [7].

The cross section of the low--energy p + p $\to$ D + ${\rm e}^+$ + $\nu_{\rm e}$ scattering is defined
\begin{eqnarray}\label{label2.18}
\sigma({\rm p} + {\rm p} \to {\rm D} + {\rm e}^+ + \nu_{\rm e})\,=\,\frac{1}{v}\,\frac{1}{4E_1E_2}\int\,\overline{|{\cal M}({\rm p} + {\rm p} \to {\rm D} + {\rm e}^+ + \nu_{\rm e})|^2}\,\times\nonumber\\
\times (2\pi)^4\,\delta^{(4)}(p_1+p_2-Q-k_{\rm e}-k_{\nu})\,\frac{d^3Q}{(2\pi)^3 2E_{\rm D}}\frac{d^3k_{\rm e}}{(2\pi)^3 2E_{\rm e}}\frac{d^3k_{\nu}}{(2\pi)^3 2E_{\nu}}\,,
\end{eqnarray}
where $\overline{|{\cal M}({\rm p} + {\rm p} \to {\rm D} + {\rm e}^+ +\nu_{e})|^2}$ is the squared amplitude averaged over polarizations of protons and summed over polarizations of final particles [7]
\begin{eqnarray}\label{label2.19}
&&\overline{|{\cal M}({\rm p} + {\rm p} \to {\rm D} + {\rm e}^+ \nu_{e})|^2}= C^2(v)\,g^2_{\rm A}M^4_{\rm N}\frac{9G^2_{\rm F}Q_{\rm D}}{16\pi^2}\times\nonumber\\
&&\times \,\frac{\displaystyle \Bigg[\frac{g^2_{\pi\rm NN}}{8M^2_{\pi}}\Bigg]^2\Bigg(1 - \frac{8\sqrt{2}\pi\,M^2_{\pi}}{g^2_{\rm \pi NN}}\,\frac{a_{\rm S}}{M_{\rm N}}\Bigg)^2}{\displaystyle 1 + \,\Bigg[\frac{g^2_{\pi\rm NN}}{8M^2_{\pi}}\Bigg]^2\Bigg(1 -  \frac{8\sqrt{2}\pi\,M^2_{\pi}}{g^2_{\rm \pi NN}}\,\frac{a_{\rm S}}{M_{\rm N}}\Bigg)^2\,\frac{M^4_{\rm N}}{4\pi^2}\,v^2}\,\Bigg(g_{\alpha\beta}-\frac{Q_{\alpha}Q_{\beta}}{M^2_{\rm D}}\Bigg)\,\times \nonumber\\
&&\times \,{\rm tr}\{(\hat{k}_{\rm e}-m_{\rm e})\gamma^{\alpha}(1-\gamma^5)\hat{k}_{\nu}\gamma^{\beta}(1-\gamma^5)\} \, \frac{1}{4}\,{\rm tr}\{(\hat{p}_1-M_{\rm N})\gamma^5 (\hat{p}_2+M_{\rm N})\gamma^5\} =\nonumber\\
&&=C^2(v)\,g^2_{\rm A}M^6_{\rm N}\frac{27G^2_{\rm F}Q_{\rm D}}{\pi^2}\,\Bigg( E_{\rm e} E_{\nu} - \frac{1}{3}\vec{k}_{\rm e}\cdot \vec{k}_{\nu}\Bigg)\,\times\nonumber\\
&&\times \,\frac{\displaystyle \Bigg[\frac{g^2_{\pi\rm NN}}{8M^2_{\pi}}\Bigg]^2\Bigg(1 - \frac{8\sqrt{2}\pi\,M^2_{\pi}}{g^2_{\rm \pi NN}}\,\frac{a_{\rm S}}{M_{\rm N}}\Bigg)^2}{\displaystyle 1 + \,\Bigg[\frac{g^2_{\pi\rm NN}}{8M^2_{\pi}}\Bigg]^2\Bigg(1 - \frac{8\sqrt{2}\pi\,M^2_{\pi}}{g^2_{\rm \pi NN}}\,\frac{a_{\rm S}}{M_{\rm N}}\Bigg)^2\,\frac{M^4_{\rm N}}{4\pi^2}\,v^2}.
\end{eqnarray}

\noindent Now we should carry out the integration over the phase volume of the final D ${\rm e}^+ \nu_{\rm e}$--state
\begin{eqnarray}\label{label2.20}
&&\int\frac{d^3Q}{(2\pi)^3 2E_{\rm D}}\frac{d^3k_{\rm e}}{(2\pi)^3 2E_{\rm e}}\frac{d^3k_{\nu}}{(2\pi)^3 2E_{\nu}}\times\nonumber\\
&&\times (2\pi)^4\,\delta^{(4)}(p_1+p_2-Q-k_{\rm e}-k_{\nu})\,\Bigg( E_{\rm e} E_{\nu} - \frac{1}{3}\vec{k}_{\rm e}\cdot \vec{k}_{\nu}\,\Bigg) = \nonumber\\
&&= \frac{1}{32\pi^2 M_{\rm N}}\,\int^{\varepsilon_{\rm D}}_{m_{\rm e}}\sqrt{E^2_{\rm e}-m^2_{\rm e}}E_{\rm e}(\varepsilon_{\rm D} - E_{\rm e})^2\,d E_{\rm e} = \frac{\varepsilon^5_{\rm D}}{960\pi^2 M_{\rm N}}\,f(\xi),
\end{eqnarray}
where $\xi=m_{\rm e}/\varepsilon_{\rm D}$ and the function $f(\xi)$ is defined by the integral
\begin{eqnarray}\label{label2.21}
f(\xi)&=&30\,\int^1_{\xi}\sqrt{x^2 -\xi^2}\,x\,(1-x)^2 dx=(1 - \frac{9}{2}\,\xi^2 - 4\,\xi^4)\,\sqrt{1-\xi^2} + \nonumber\\
&&+ \frac{15}{2}\,\xi^4\,{\ell n}\Bigg(\frac{1+\sqrt{1-\xi^2}}{\xi}\Bigg) = 0.776 
\end{eqnarray}
and normalized to unity at $\xi=0$, i.e., $f(0)=1$.

The cross section of the low--energy two--proton fusion p + p $\to$ D + ${\rm e}^+$ + $\nu_{\rm e}$ is given by 
\begin{eqnarray}\label{label2.22}
&&\sigma({\rm p} + {\rm p} \to {\rm D} + {\rm e}^+ + \nu_{\rm e}) =
\frac{C^2(v)}{v}\,\frac{9g^2_{\rm A}G^2_{\rm F}Q_{\rm D}}{1280 \pi^5}\,\,\varepsilon^5_{\rm D}\,M^3_{\rm N}f\Bigg(\frac{m_{\rm e}}{\varepsilon_{\rm D}}\Bigg) \times \nonumber\\
&&\times\,\frac{\displaystyle \Bigg[\frac{g^2_{\pi\rm NN}}{8M^2_{\pi}}\Bigg]^2\Bigg(1 - \frac{8\sqrt{2}\pi\,M^2_{\pi}}{g^2_{\rm \pi NN}}\,\frac{a_{\rm S}}{M_{\rm N}}\Bigg)^2}{\displaystyle 1 +\,\Bigg[\frac{g^2_{\pi\rm NN}}{8M^2_{\pi}}\Bigg]^2\Bigg(1 - \frac{8\sqrt{2}\pi\,M^2_{\pi}}{g^2_{\rm \pi NN}}\,\frac{a_{\rm S}}{M_{\rm N}}\Bigg)^2 \frac{M^4_{\rm N}}{4\pi^2}\,v^2}=\nonumber\\
&&= \frac{1}{\displaystyle 1 + 52743\,v^2}\,\times\,\frac{C^2(v)}{v}\,\times\,1.34\times10^{-48}\,{\rm cm}^2.
\end{eqnarray}
The cross section is calculated in units of $\hbar=c=1$. The appearance of the factor $C^2(v)$ agrees with the result obtained by Bethe and Critchfield [4].

The reaction rate of the process p + p $\to$ D + ${\rm e}^+$ + $\nu_{\rm e}$ reads [7]
\begin{eqnarray}\label{label2.23}
&&<v\,\sigma({\rm p} + {\rm p} \to {\rm D} + {\rm e}^+ + \nu_{\rm e})> = 1.03\times 10^{-39}\,\times\,\frac{1}{\alpha}\,\times\,\frac{2}{\pi}\times\,\Bigg(\frac{1}{3}\Bigg)^2\times\nonumber\\
&&\times\,\sqrt{\frac{\pi}{3}}\,\times\,\frac{\displaystyle \tau^2\,e^{-\tau}}{\displaystyle 1 + 52743\,\frac{3\alpha^2 \pi^2}{\tau}} = 1.03\times 10^{-38}\,\frac{\displaystyle \tau^2\,e^{-\tau}}{\displaystyle 1 + \frac{83}{\tau}}\,\;{\rm cm}^3\,{\rm s}^{-1},
\end{eqnarray}
where $\tau$ is connected with the temperature $T$ [4]
\begin{eqnarray}\label{label2.24}
\tau = 3\Bigg(\frac{\alpha^2\pi^2 M_N}{4 kT}\Bigg)^{1/3}.
\end{eqnarray}
and $k\,=\,8.62\times 10^{-11}\,{\rm MeV}\cdot {\rm K}^{-1}$ is the Boltzmann constant. Following [4] we have computed the integral over relative velocities of the protons by a saddle point approximation.

Setting $T = T_{\rm c} = 15.5$, measured in unites of $10^6\,{\rm K}$, where  $T_{\rm c}$ is the temperature of the solar core in the Standard Solar model [1,2], we obtain $\tau=13.56$ and the reaction rate is given by
\begin{eqnarray}\label{label2.25}
<v\,\sigma({\rm p} + {\rm p} \to {\rm D} + {\rm e}^+ + \nu_{\rm e})>\, =\,3.44\times 10^{-43}\,{\rm cm}^3\,{\rm s}^{-1}\,.
\end{eqnarray}
This value differs by a factor of 2.90$\pm$0.87 from that calculated within the potential approach [1,2] (see also Ref.~[5]).
In order to reconcile our result with the solar luminosity we must assume that the temperature in the solar core equals $T_{\rm c}= 13.8^{-0.4}_{+0.5} \times 10^6\,{\rm K}$ [3]
here we assumed. As has been remarked in Ref.~[3] the increase of the magnitude of the cross section of the two--proton fusion leads to a decrease of the temperature in the solar core, and  provides a strong suppression of the solar neutrino flux. 

The enhancement of the amplitude of the p + p $\to$ D + W transition found in our approach is related to the computation of the amplitude in terms of one--nucleon loop diagrams. Indeed, the structure function Eq.~(\ref{label2.10}) defining the effective Lagrangian of the p + p $\to$ D + W transition is due to the contribution of the anomalous part of the AAV one--nucleon loop diagram [11,13,15]. Such an anomalous contribution, produced by vacuum fluctuations of virtual nucleons, has a quantum field theory nature and cannot be obtained within the potential approach describing strong low--energy interactions of the protons and the deuteron in terms of the overlap integral of the wave--functions of two--protons and the deuteron. The ambiguity of the computation of the AAV--anomaly, produced by the shift of virtual nucleon momentum, has been fixed by the requirement of gauge invariance under gauge transformations of the deuteron field [7]. This is very similar to the removal of the ambiguity appearing for the computation of the Adler--Bell--Jackiw--Bardeen anomaly [20].

\section{Solar neutrino fluxes}
\setcounter{equation}{0}

\hspace{0.2in} The solar neutrino signals ${\rm S}_{\rm Cl}$ and ${\rm S}_{\rm Ga}$ obtained in the Gallium [21] and Chlorine [22] experiments, respectively, are given by

\parbox{11cm}{\begin{eqnarray*}
{\rm S}_{\rm Ga}(E_{\nu}\ge 0.233\;{\rm MeV})&=&(77 \pm 8)\;{\rm SNU},\\
{\rm S}_{\rm Cl}(E_{\nu}\ge 0.814\;{\rm MeV})&=&(2.55 \pm 0.25)\;{\rm SNU},\\
{\rm S}_{\rm Kam}(E_{\nu}\ge 7\;{\rm MeV})&=&(2.73\pm 0.38) \cdot 10^6\;\;{\rm cm}^{-2}{\rm s}^{-1},
\end{eqnarray*}} \hfill
\parbox{1cm}{\begin{eqnarray}\label{label3.1}
\end{eqnarray}}

\noindent where the Solar Neutrino Unit (SNU) is defined as one reaction per second per $10^{36}$ atoms [3]. We have adduced also the data given by KAMIOKANDE experiment [23] (see also Ref.~[3]). Due to the energy threshold in KAMIOKANDE experiment this neutrino detector is only sensitive to ${^8}{\rm B}$ neutrinos.

The solar neutrino fluxes $\Phi_i$, where i = pp, pep,$^7$Be, $^{13}$N, $^{15}$O and $^8$B, etc. can be represented in the form of a power--law behavior (see Eqs.~(96) in Ref.~[3]), i.e.,
\begin{eqnarray}\label{label3.2}
\Phi_i\,=\,x^{\alpha_i}\,\Phi^*_i.
\end{eqnarray}
$\Phi^*_i$ is the neutrino flux, predicted by SSM, and the parameter x  in our definition reads
\begin{eqnarray}\label{label3.3}
x\,=\,\frac{<v\,\sigma({\rm p} + {\rm p} \to {\rm D} + {\rm e}^+ + \nu_{\rm e})_{\rm S}>}{<v\,\sigma({\rm p} + {\rm p} \to {\rm D} + {\rm e}^+ + \nu_{\rm e})_{\rm S}>^*}=2.90 \pm 0.87,
\end{eqnarray}
where $<v\,\sigma({\rm p} + {\rm p} \to {\rm D} + {\rm e}^+ + \nu_{\rm e})_{\rm S}>^*$ is the quantity calculated in the potential approach.

The values of the parameters ${\alpha}_{\rm i}$ are given in Table~1 and can be found in Table X of Ref.~[3]. In Table~1 we also adduce the neutrino fluxes that should give the contributions to the signals detected in Gallium and Chlorine experiments. We have normalized our predictions to the results obtained within SSM [2].

It is seen that our model explains reasonably well the experimental data of the Gallium experiment. For the neutrino flux measured in the Chlorine experiment our model leaves room for possible other contributions. One can see that our prediction for ${\rm S}_{\rm Kam}(E_{\nu}\ge 7\;{\rm MeV})=\Phi_{\rm B}[10^6\;{\rm cm}^{-2}{\rm s}^{-1}]=0.37^{-0.19}_{+0.61}$ (see Table IX of Ref.~[3]) is too small when compared with the KAMIOKANDE experimental data.

\section{Conclusion}

\hspace{0.2in} We resume the obtained results. We have shown that the relativistic field theory model of the deuteron, describing the strong interactions of the deuteron with hadrons in terms of one--nucleon loop diagrams, has provided contributions of strong interactions to the matrix element of two--proton fusion p + p $\to$ D + ${\rm e}^+$ + $\nu_{\rm e}$, that has led to an enhancement of the reaction rate by a factor 2.9 with respect to the computations performed within non--relativistic approaches. 

This explains  reasonably well the experimental data of the Gallium experiment. For the neutrino flux measured in the Chlorine and KAMIOKANDE experiment we obtain smaller values. This implies that the enhancement of the cross section of the two--proton fusion obtained in our model leads to a suppression of the high energy neutrino 
fluxes. 

In the conclusion let us discuss the estimate of the theoretical uncertainty of the relativistic field theory model of the deuteron. Unfortunately, in our model there is not any small parameter like $\alpha=e^2/4\pi=1/137$, the fine structure constant in QED or a large parameter $N$, the number of quark colors in a multicolor extension of QCD with $SU(3)_c\to SU(N)_c$, that allows to apply the expansion in powers of $1/N$, so--called large N--expansion, to quark--gluon diagrams describing strong low--energy interactions of hadrons.

Our effective approach is based on the one--nucleon loop diagram approximation for the description of both the self--interactions of the deuteron and coupling of the deuteron to other particles. By describing the self--interactions of the deuteron at the one--nucleon loop approximation we have fitted all parameters characterizing the physical deuteron. In this case the only way to estimate the theoretical uncertainty of the approach is to compared the one--nucleon loop data with the two--nucleon loop corrections. In Refs.~[6,7] we have calculated the two--nucleon loop correction to the binding energy of the deuteron. This correction reads [6,7]
\begin{eqnarray*}
(\delta\,\varepsilon_{\rm D})_{\rm two-loop}=\Lambda_{\rm D}\,\frac{11}{3}\,\frac{g^2_{\pi{\rm NN}}}{4\pi}\,\frac{g^2_{\rm V}}{3\pi^3}\Bigg(\frac{\Lambda_{\rm D}}{M_{\pi}}\Bigg)^2 \Bigg(\frac{\Lambda_{\rm D}}{M_{\rm N}}\Bigg)^3 = 0.36\,{\rm MeV}.
\end{eqnarray*}
The numerical value of the two--nucleon loop correction makes up about 16$\%$ of the binding energy $\varepsilon_{\rm D}=2.273\,{\rm MeV}$ calculated in the one--nucleon loop approximation [7]. Thus the magnitude $\Delta =\pm 16\,\%$ might be accepted as a theoretical uncertainty of the relativistic field theory model of the deuteron for amplitudes. The theoretical uncertainty of the cross section should be about 30$\,\%$. This agrees well with our prediction for the cross section of the radiative neutron--proton capture. Indeed, the theoretical value $\sigma({\rm n} + {\rm p} \to {\rm D} + \gamma) = 276\,{\rm mb}$ [7] is in agreement with the experimental data $\sigma({\rm n} + {\rm p} \to {\rm D} + \gamma)_{\exp}\,=\,(334.2 \pm 0.5)\,{\rm mb}$ with an accuracy of 22$\,\%$. By taking into account the theoretical uncertainty, the cross section of the radiative neutron--proton capture should read: $\sigma({\rm n} + {\rm p} \to {\rm D} + \gamma) = (276\pm 83)\,{\rm mb}$. Taking into account the theoretical uncertainty of the model we predict for the two--proton fusion reaction rate  an enhancement by a factor of $2.90\pm 0.87$ compared to the potential approach.

\section{Acknowledgement}

\hspace{0.2in} This work was partially supported  by the Fonds zur F\"orderung der wissenschaftlichen Forschung in \"Osterreich (project P10361--PHY). Discussions with Prof.~G.~E.~Rutkovsky are appreciated.

\newpage
\begin{center}
\section{References}
\end{center}

\begin{description}
\item{[1]}~Rolfs~C.~E. $\&$ Rodney~W.~S. 1988, in Cauldrons in Cosmos, The University of Chicago Press (Chicago and London) and references therein.
\item{[2]}~Bachall~J.~N. 1989, in Neutrino Astrophysics, Cambridge University Press (Cambridge -- New York -- New Rochelle--Melbourne--Sydney) 60;
\item{~~~}~Bahcall~J.~N. $\&$ Pinsonneault M.~H.~1995, Rev. Mod. Phys., 67, 781.
\item{[3]}~Castellani~V., Degl$^{\prime}$Innocenti  S., Fiorentini G. , Lissia  M. $\&$ Ricci B. 1997, Phys. Rep., 281, 309.
\item{[4]}~Bethe~H.~A. $\&$ Critchfield C.~L.~1939, Phys. Rev., 54, 248.
\item{[5]}~Kamionkowski~M. $\&$ Bahcall J.~N.~1991, ApJ., 359, 884;
\item{~~~}~Bahcall~J.~N. $\&$ Pinsonneault M.~H.~1992, Rev. Mod. Phys., 64, 885.
\item{[6]}~Ivanov~A.~N.,~Troitskaya~N.~I., Faber~M.~ $\&$ Oberhummer~H.~1995, Phys. Lett. B361, 74.
\item{[7]}~Ivanov~A.~N.,~Troitskaya~N.~I., Faber~M.~ $\&$ Oberhummer~H.~1996, On the relativistic field theory model of the deuteron II, Nucl. Phys. A (in press) and Erratum (submitted to Nucl. Phys. A);
nucl--th/9704031.
\item{[8]}~Adler~S.~L.~1969, Phys. Rev., 177, 2726.
\item{[9]}~Bell~J.~S. $\&$ Jackiw~R.~1969, Nuovo Cim., 60A, 47;
\item{~~~}~Bardeen~W.~A.~1969, Phys. Rev., 182, 1517.
\item{[10]}~Wilson~K.~G., Phys.~Rev., 181, 1909.
\item{[11]}~Gertsein~I.~S. $\&$ Jackiw~R.~1969, Phys.~Rev.,181, 1955.
\item{[12]}~Brown~R.~W.,~Shih~C.-C. $\&$ ~Young~B.-L.1969, Phys.~Rev., 186, 1491.
\item{[13]}~Wess~J. $\&$ Zumino~B.~1971, Phys. Lett., B37, 95.
\item{[14]}~Ivanov ~A.~N. $\&$ Shechter~V.~M.~1980, Sov. J. Nucl. Phys., 31,  275 and references therein.
\item{[15]}~Ivanov~A.~N.~1981,~Sov. J. Nucl. Phys., 33, 904. 
\item{[16]}~Nagels~M.~M. et al.~1979, Nucl. Phys., B147, 253.
\item{[17]}~Hornyak~W.~F. 1975, in Nuclear Structure, Academic Press ( New York -- San Francisco -- London), 495.
\item{[18]}~Itzykson~C. $\&$ Zuber J.-B.~1980, in Quantum Field Theory (McDraw--Hill),~6--14.
\item{[19]}~Scadron~M.~D.~1979, in Advanced in Quantum Theory and its Applications Through Feynman Diagrams, Springer--Verlag (New York--Heidelberg), 237--239.
\item{[20]}~Jackiw~R.~1972, in Lectures on Current Algebra and its applications, Princeton (University Press). 
\item{[21]}~Devis ~R.~Jr.,~Harmer~D.~S.$\&$ Hoffman~K.~C.~1968, Phys. Rev. Lett., 20, 1205.
\item{~~~}~Cleveland ~B.~T. et al.~1995, Nucl. Phys., B38 (Proc. Suppl.), 47 and references therein.
\item{[22]}~GALLEX Collaboration, Anselmann ~P. et al.~1994, Phys. Lett. B327, 377;
\item{~~~}~GALLEX Collaboration, Anselmann ~P. et al.~1995, Phys. Lett. B357, 237 and references therein;
\item{~~~}~Abazov ~A.~I. et al.~1991, Phys. Rev. Lett., 67, 3332;
\item{~~~}~SAGE Collaboration,  Abdurashitov~J.~N. et al.~1996, Nucl. Phys., B48 (Proc. Suppl.), 299 and references therein.
\item{[23]}~Hirata~K.~S.~et al.~1989, Phys. Rev. Lett., 63, 16;
\item{~~~}~Kajita~T, ICRR--Report, 332--94--27 (December 1994), and references therein.
\end{description}

\newpage

\noindent Table 1.  Contributions from the main components of the neutrino fluxes to the signals (SNU) in the Gallium [21] and Chlorine [22] experiments according to the SSM [2] and our approach (see Table VII of Ref.~[3]). The errors are due to the assumed 30\,\% uncertainty of the reaction rate for p + p $\to$ D + $e^{\,+}$ + $\nu_{\rm e}$.
The power--law parameters $\alpha_i$ have been taken from Table X of Ref.~[3].
\vspace{0.2in}

\begin{tabular}{|c||c||c|c||c||c|c||c|  }\hline
 & \multicolumn{3}{c||}{S$_{\rm Ga}(E_{\nu}\ge 0.233\;{\rm MeV})$}& \multicolumn{3}{c||}{S$_{\rm Cl}(E_{\nu}\ge 0.814\;{\rm MeV})$}& \\ \cline{2-7}
& SSM & our model& experiment & SSM & our model & experiment & \raisebox{1.5ex}[-1.5ex]{$\alpha_i$}  \\ \hline
pp & 69.7 & 75.1$^{+1.1}_{-1.9}$ &  & 0.00 & 0.00 & &0.07 \\
pep & 3.0 & 3.2$^{+0,0}_{-0.0}$ &  & 0.22 & 0.24$^{+0.01}_{-0.01}$ & & 0.07 \\
$^7{\rm Be}$ & 37.7 & 11.7$^{-2,9}_{+5.6}$ &  & 1.24 & 0.39$^{-0.10}_{+0.17}$ & &$-1.1$ \\
$^{13}{\rm N}$ & 3.8 & 0.4$^{-0.2}_{+0.4}$ &  & 0.11 & 0.01$^{-0.01}_{+0.01}$ & &$-2.2$ \\
$^{15}{\rm O}$ & 6.3 & 0.6$^{-0.3}_{+0.6}$ &  & 0.37 & 0.04$^{-0.01}_{+0.04}$ & &$-2.2$ \\
$^{8}{\rm B}$ & 16.1 & 0.9$^{-0.4}_{+1.5}$ &  & 7.36 &  0.42$^{-0.22}_{+0.67}$ & &$-2.7$ \\ \hline
 & 136.6 & 91.9$^{-2.9}_{+6.2}$ & $77.1 \pm 13.4$ & 9.30 & 1.10$^{-0.33}_{+0.88}$ & $2.55\pm 0.35$ &\\ \hline
\end{tabular}\\

\newpage

\section*{\bf Figure Captions}
\begin{itemize}
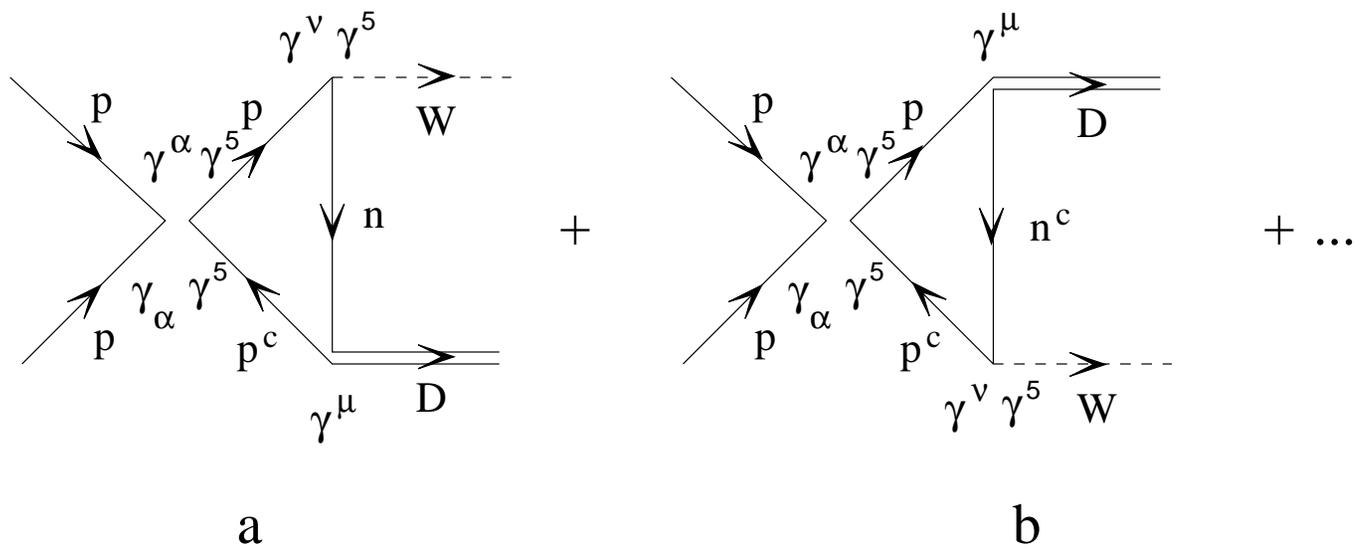
\item{Fig.~1}
The contribution of $[\bar{p}(x)\gamma_{\mu}\gamma^5 p^c(x)][\bar{p^c}(x)\gamma^{\mu}\gamma^5 p(x)]$ to the amplitude of the p + p $\to$ D + ${\rm e}^+$ + $\nu_{\rm e}$ scattering.
\end{itemize}

\end{document}